\documentclass[twocolumn,preprintnumbers, amssymb,amsmath,aps,floatfix,prd,nofootinbib,superscriptaddress,showpacs]{revtex4-1}

\usepackage{epsfig}
\usepackage{bm}
\usepackage{amssymb}
\usepackage{amsmath}
\usepackage{color}
\usepackage{subfigure}
\usepackage[colorlinks,
            linkcolor=blue,
            anchorcolor=black,
            citecolor=blue
            ]{hyperref}

\begin{document}

\title{Momentum Anisotropy of Leptons from Two Photon Processes in Heavy Ion Collisions}

\author{Bo-Wen Xiao}
\affiliation{Key Laboratory of Quark and Lepton Physics (MOE) and Institute
of Particle Physics, Central China Normal University, Wuhan 430079, China}
\affiliation{School of Science and Engineering, The Chinese University of Hong Kong, Shenzhen 518172, China}

\author{Feng Yuan}
\affiliation{Nuclear Science Division, Lawrence Berkeley National
Laboratory, Berkeley, CA 94720, USA}

\author{Jian Zhou}
\affiliation{\normalsize\it  Key Laboratory of Particle Physics and
Particle Irradiation (MOE),Institute of Frontier and
Interdisciplinary Science, Shandong University (QingDao), Shandong
266237, China }


\begin{abstract}
We investigate the azimuthal angular correlation between the lepton transverse momentum $P_\perp$ and the impact parameter $b_\perp$ in non-central heavy-ion collisions, where the leptons are produced through two-photon scattering. Among the Fourier harmonic coefficients, a significant $v_4$ asymmetry is found for the typical kinematics at RHIC and LHC with a mild dependence on the $P_\perp$, whereas $v_2$ is power suppressed by the lepton mass over $P_\perp$. This unique prediction, if confirmed from the experiments, shall provide crucial information on the production mechanism for the dilepton in two-photon processes.
\end{abstract}
\maketitle

\section{Introduction}
The flow phenomenon of final state particles in heavy-ion collisions is one of the most important observations that signals the collective modes of the quark-gluon plasma created in these collisions~\cite{Back:2004je,Adams:2005dq,Adcox:2004mh,Aamodt:2010pa,ATLAS:2012at,Chatrchyan:2013nka}. They are defined as the anisotropy of final state hadrons in the transverse plane with respect to the impact parameter of the collision~\cite{Ollitrault:1992bk}, e.g., in terms of $\cos(n\phi)$ where $\phi$ is the azimuthal angle between the hadron's momentum $\vec{p}_{h\perp}$ and the impact parameter $\vec{b}_\perp$. In this paper, we study the momentum anisotropy of the leptons from the pure electromagnetic process of $\gamma\gamma\to \ell^+\ell^-$ in heavy ion collisions. This anisotropy refers to the angular distribution of the leptons with respect to the reaction plane defined by the impact parameter of the collision. Although the associated phenomena may strongly resemble the conventional hadronic flow in experimental measurements, its underlying physics mechanism is from the initial state interactions. The comparison of the anisotropy between the leptons and hadrons will provide a unique perspective for the anisotropy phenomena in heavy-ion collisions.

Di-lepton production through the QED processes in heavy-ion collisions has a long history, mainly in the so-called Ultra-Peripheral Collisions~(UPC)~\cite{Bertulani:1987tz,Adams:2004rz,Baur:2003ar,Hencken:2004td,Baur:2007fv,Bertulani:2005ru,{Baltz:2007kq},Baltz:2009jk,ATLAS:2016vdy,Klein:2018cjh,CMS:2020avp}. More recently, experiments at RHIC and LHC have pushed these measurements toward peripheral and central collisions. This was achieved by selecting the dilepton events through the kinematic constraints where the total transverse momentum of the dilepton is very small, well below the typical hadronic scale of $0.3~\rm GeV$. Significant deviations from the UPC case have been reported~\cite{Aaboud:2018eph,Adam:2018tdm,Lehner:2019amb,Adam:2019mby,ATLAS:2019vxg}, where the mean value of the total transverse momentum of the lepton pair $q_\perp$ increases with centrality. Especially, from the ATLAS measurement, it reaches a value of about $100~\rm MeV$ in the most central collisions at the LHC, whereas it is about $40~\rm MeV$ for UPC case~\cite{Klein:2018fmp}. In this particular kinematic region, the lepton pairs are predominantly produced by the coherent electromagnetic fields of the incoming nuclei as initially demonstrated in Ref.~\cite{Klusek-Gawenda:2018zfz}.  These developments have generated quite an interest in the heavy-ion community. If it is confirmed that the observed effects indeed come from the medium interactions with the lepton pair, this shall lead to a potential probe to the electromagnetic property of the hot medium~\cite{Aaboud:2018eph,Adam:2018tdm,Klein:2018fmp}. Therefore, the key step is to quantify the initial state contributions from the two-photon processes. To do that, we have to go beyond the previous calculations which only apply to the dilepton production in UPC events~\cite{Klein:2018fmp,Zha:2018tlq,Li:2019yzy,Li:2019sin,Zhao:2019hta,Karadag:2019gvc}.

On the other hand, this extension is not straightforward, since we have to compute the joint transverse momentum and impact parameter dependence for the incoming photon fluxes of the colliding nuclei~\cite{Vidovic:1992ik,Klein:2020jom}. Different assumptions and models have been introduced~\cite{Vidovic:1992ik,Klein:2018fmp,Zha:2018tlq,Li:2019yzy,Li:2019sin,Zhao:2019hta,Karadag:2019gvc,Klein:2020jom}. Among these calculations, the so-called QED calculation seems to suggest that the observed $P_T$-broadening effects may solely come from the initial state effects due to different geometry of the collisions~\cite{Zha:2018tlq}. However, the predicted azimuthal $\cos(2\phi)$ asymmetry between the total transverse momentum $q_\perp$ and the impact parameter $b_\perp$ remains to be confirmed in experiments~\cite{RuanXu}. This asymmetry depends on simultaneously determining the transverse momentum and impact parameter information and needs further studies.

The proposed anisotropy measurement in this paper is different from those in Refs.~\cite{Zha:2018tlq,Li:2019yzy,Li:2019sin}, where the azimuthal asymmetry depends on the total transverse momentum of the lepton pair $q_\perp$. In our study, the anisotropy is defined through the lepton momentum $P_\perp$ with respect to the impact parameter $b_\perp$. Therefore, we can integrate over $q_\perp$ to a certain value, e.g.,  $100~\rm MeV$, where the electromagnetic contribution dominates over the hadronic contributions as demonstrated in experiments at RHIC and LHC~\cite{Aaboud:2018eph,Adam:2018tdm,Lehner:2019amb,Adam:2019mby,ATLAS:2019vxg}. Because the total transverse momentum $q_\perp$ (order $100\rm MeV$) is so small compared to the lepton momentum (order $\rm GeV$), the lepton pair are almost back-to-back in the transverse plane. 

In our calculations, the photon fluxes only depend on the impact parameter $b_\perp$ and can be rigorously computed through the classic electromagnetic treatments, like the Jackson method~\cite{Jackson:1998nia}. The impact parameter dependent photon flux predicts a significant linear polarization along the impact parameter direction. This will generate a $\cos(4\phi)$ azimuthal asymmetries between the lepton's transverse momentum and the impact parameter $b_\perp$, resulting in an anisotropy. 

The rest of our paper is organized as follows. In Sec.~II, we briefly discuss the impact parameter dependent photon flux and its polarization for a moving ion. In Sec.~III, we derive the lepton anisotropy in the two-photon process due to the incoming photons' polarizations. Numeric results will be shown for RHIC and LHC experiments in the relevant kinematics. We conclude our paper in Sec.~IV.

\section{Polarization in Incoming Photon Flux}

\begin{figure}[tbp]
\begin{center}
\includegraphics[width=6.6cm]{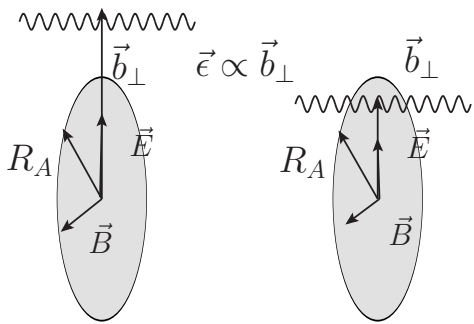}
\end{center}
\caption[*]{Illustration of the polarized photon flux associated with a relativistic heavy nucleus moving to the right. The physical polarization of the photon propagating to the right is along the direction of $\vec{b}_{1\perp}$ respect to the center of the nucleus in the transverse plane.}
\label{polarizationb}
\end{figure}

When a heavy ion moves at an ultra-relativistic speed, e.g., along the $\hat z$ direction, it coherently generates associated electromagnetic (EM) fields when the wavelength of the EM field is comparable with nucleus radius. These EM fields can be described as an effective photon flux~\cite{Fermi:1924tc,vonWeizsacker:1934nji,Williams:1934ad}. As illustrated in Fig.~\ref{polarizationb}, not only the intensity but also the polarization of the photon flux depend on the impact parameter $\vec{b}_{1\perp}$, where $b_{1\perp}$ represents the transverse distance relative to the center of the moving nucleus. 

From the perspective of the classical electrodynamics, as pointed out in Refs.~\cite{Jackson:1998nia, Jackiw:1991ck} and shown in Fig.~\ref{polarizationb}, the electric field $\vec E$ generated by the relativistically moving charged nucleus is linearly polarized along with the impact parameter $\vec b_{1\perp}$ direction, and the corresponding magnetic field $\vec B$ is perpendicular to the electric field in the transverse plane. Therefore, a physical gauge choice of the polarization vector $\vec \epsilon_\perp = \hat b_{1\perp}$ with $\hat b_{1\perp}$ the unit vector of $\vec b_{1\perp}$ can be used to a quantum field theory calculation, especially when the polarization of the equivalent photon plays an important role.

The above physics is appropriately captured by introducing the photon distribution from the nucleus. Following the example of the generalized parton distribution (GPD) in nucleon~\cite{Ji:1996ek}, we introduce the generalized photon distribution. Similar to that of quark/gluon GPD~\cite{Burkardt:2000za}, the photon GPD can be interpreted as the impact parameter dependent photon distribution. This is equivalent to the photon flux discussed in the literature. The photon GPD is defined through the following matrix,
\begin{eqnarray}
   && xf_\gamma^{\alpha\beta}(x;b_{1\perp})=\int\frac{d^2k_\perp d^2\Delta_\perp}{(2\pi)^2}e^{i\Delta_\perp\cdot b_{1\perp}}\int\frac{d^2r_\perp}{(2\pi)^2}e^{ik_\perp\cdot r_\perp}\nonumber\\
    &&~~\times \langle A,-\frac{\Delta_\perp}{2}|F^{+\alpha}(\frac{r_\perp}{2})F^{+\beta}(-\frac{r_\perp}{2})|A,\frac{\Delta_\perp}{2}\rangle \ ,
\end{eqnarray}
where $F^{\mu\nu}$ represent the EM field strength. The photon GPD can be parameterized as
\begin{eqnarray}
    xf_\gamma^{\alpha\beta}(x;b_{1\perp})&=&\frac{\delta^{\alpha\beta}}{2}xf_\gamma(x;b_{1\perp})\nonumber\\
   && +\left(\frac{b_{1\perp}^\alpha b_{1\perp}^\beta}{b_{1\perp}^2}-\frac{\delta^{\alpha\beta}}{2}\right)xh_\gamma(x;b_{1\perp}) \ .
\end{eqnarray}
Here, $ f_\gamma(x, b_{1\perp})$ is the normal polarization averaged impact parameter dependent photon distribution, and $h_\gamma(x,b_{1\perp} )$ is conventionally referred to as the helicity flip photon GPD, similar to the helicity flip gluon GPD~\cite{Hoodbhoy:1998vm,Belitsky:2000jk,Diehl:2001pm}. When $x$ is sufficiently small ($x<\frac{1}{R_A m_p}$), photon distribution is dominated by these coherently generated due to the $Z^2$ enhancement where $Z$ is the nuclear charge number. By treating the external electromagnetic field of a relativistic nucleus as a classical Coulomb potential, the associated coherent photon distributions can be readily computed in terms of the nuclear charge form factor,
\begin{eqnarray}
 xh_\gamma(x, b_{1\perp})&=&xf_\gamma(x,b_{1\perp})\nonumber\\ 
&=&4Z^2\alpha
\left|\int\frac{d^2{q_\perp}}{(2\pi)^2}e^{iq_\perp\cdot b_{1\perp}}\frac{\vec{q}_\perp}{q^2}F_A(q^2)\right|^2 \ ,\label{fluxb1}
\end{eqnarray}
where $q^2=q_\perp^2+x^2m_p^2$ and $F_A$ represents the EM form factor for the nucleus, and $m_p$ being proton mass. One finds that for a given $b_{1\perp}$, coherent photons are fully linearly polarized due to the fact that $h_\gamma=f_\gamma$ in the above equation, see also the discussions in Ref.~\cite{Baur:2003ar}. This relation essentially is the consequence of the property of highly boosted Coulomb field: the direction of the electric field generated by a spherically symmetric charge source distribution is parallel to the impact parameter. The similar relation between unpolarized photon TMD and linearly polarized photon TMD was also established in Ref.~\cite{Li:2019yzy}. In Ref.~\cite{Zha:2018ywo}, the above photon flux $f_\gamma(x,b_{1\perp})$ has been applied to understand the dilepton production in peripheral collisions, and it was found that the photon flux at the small impact parameter of $b_\perp<R_A$ plays a significant role in the non-UPC events in heavy-ion collisions. In the following, we will derive the lepton's anisotropy from the helicity flip photon GPD $h_\gamma(x,b_{1\perp})$.

\section{Anisotropy of Leptons in Two-photon Processes}

One naturally expects that the helicity flip photon GPD could introduce a $\cos 4 \phi$ modulation in the azimuthal distribution of di-lepton produced in two-photon processes as the linearly polarized photon TMD does due to the similar  photon polarization tensor structure. However, the $\cos 4 \phi$ azimuthal asymmetries induced by the helicity flip photon GPD and the linearly polarized photon TMD are different types. The angle $\phi$ here refers to the azimuthal angle between leading lepton transverse momentum and the impact parameter of heavy-ion collisions. On the other hand, the $\cos 4 \phi$ azimuthal asymmetry investigated in the previous work~\cite{Li:2019yzy,Li:2019sin} describes the correlation between lepton transverse momentum and the total transverse momentum of lepton pair.

The dominant channel for di-lepton production in peripheral and ultraperipheral heavy-ion collisions are the Breit-Wheeler process $\gamma(x_1P)+\gamma(x_2 \bar P) \rightarrow \ell^+(l_1)+\ell^-(l_2)$ where leptons are produced nearly back-to-back in the transverse plane and the total transverse momentum $q_\perp$ is small (order $100\rm MeV$).  For convenience, we further define $\vec{P}_\perp=(\vec{l}_{1\perp}-\vec{l}_{2\perp})/2$, where $\vec{l}_{1\perp}$ and $\vec{l}_{2\perp}$ are the transverse momenta for the final state two leptons. In the back-to-back configuration, $P_\perp$ is approximately equal to the leading lepton's transverse momentum. 

The differential cross section for the lepton will be normally azimuthal angular symmetric. However, the helicity-flip photon will contribute to an azimuthal asymmetry. The reason is the following. As shown in the previous section, the photon polarizations are correlated to the individual impact parameters $b_{1\perp}$ and $b_{2\perp}$ of the nuclei. Here, $\vec{b}_{1,2\perp}$ represent the positions of two incoming nuclei with respect to the interaction point where the lepton pair is produced. The collision impact parameter $b_\perp$, which is the distance between two colliding nuclei centers, can be written as $\vec{b}_\perp=\vec{b}_{1\perp}-\vec{b}_{2\perp}$. For a particular centrality bin with the corresponding impact parameter $b_\perp$, we integrate out $b_{1\perp}$ and $b_{2\perp}$ with this constraint. Following Ref.~\cite{Li:2019yzy}, the azimuthal angle dependence can be computed from the lowest order QED which gives the following amplitude square
\begin{equation}
\left| \mathcal{M}\right|^2 = 2e^4 \left[ \left(\frac{u}{t}+\frac{t}{u}\right)-2 \cos \left(2\left(\phi_1 +\phi_2\right)\right) \right]\ , \label{eq:e4}
\end{equation}
with $u$ and $t$ being the usual Mandelstam variables, the azimuthal angles $\phi_1$ and $\phi_2$ being the angles of $\vec b_{1\perp}$ and $\vec b_{2\perp}$ with respect to $\vec{P}_\perp$.
To obtain the differential cross section depending on the collision impact parameter $\vec{b}_\perp$, we integrate out $\phi_1$ and $\phi_2$ of the above equation with the constraint that $\vec{b}_\perp=\vec{b}_{1\perp}-\vec{b}_{2\perp}$. After the integration, the $\phi_1$ and $\phi_2$ dependence in Eq.~(\ref{eq:e4}) will convert into a $\phi$-dependence where $\phi$ is the azimuthal angle between $\vec{b}_\perp$ and $\vec{P}_\perp$. In particular, the first term leads to an isotropic distribution of $\phi$, whereas the second term $-2\cos(2(\phi_1+\phi_2))$ results into an anisotropy of $\cos(4\phi)$. Therefore, in addition to the usual isotropic term in the cross section, there is a nonzero $v_4$ in the leading contributions as follows
\begin{eqnarray}
\frac{d\sigma}{d^2 P_{\perp} dy_1 dy_2 d^2 b_\perp }=
 \frac{2\alpha_e^2}{Q^4} \left [ \mathcal{A}+\mathcal{C} \cos 4\phi \right ] \ , \label{eq:e5}
\end{eqnarray}
where $y_1$ and $y_2$ are leptons' rapidities, respectively. $Q^2=x_1 x_2 s$ is the invariant mass square of the lepton pair, where the incoming photons longitudinal momentum fractions $x_1=\sqrt{{P_\perp^2}/{s}}(e^{y_1}+e^{y_2})$ and $x_2=\sqrt{{P_\perp^2}/{s}}(e^{-y_1}+e^{-y_2})$ are determined by the external kinematics. Here we have neglected the lepton mass dependence in $x_{1,2}$ for the typical kinematics at RHIC and the LHC. The coefficients $\mathcal{A}$ and $\mathcal{C}$ read
\begin{eqnarray}
\mathcal{A}&=&  \frac{Q^2-2 P_\perp^2}{P_\perp^2} \int d^2b_{1\perp} d^2 b_{2\perp} \delta^{(2)}( \vec b_\perp-\vec b_{1\perp}+\vec b_{2\perp}) \nonumber\\
&&~~\times x_1f_\gamma(x_1, b_{1\perp}^2) x_2f_\gamma(x_2, b_{2\perp}^2) \label{ps} \ ,\\
\mathcal{C} &=& -2 \int d^2b_{1\perp} d^2 b_{2\perp}
 \delta^{(2)}( \vec b_\perp-\vec b_{1\perp}+\vec b_{2\perp}) \nonumber \\
&&~~\times \left [2\left (\! 2(\hat b_{2\perp} \! \cdot \hat b_{\perp})
(\hat b_{1\perp} \! \cdot \hat b_{\perp}) -\hat b_{1\perp} \!\cdot \!\hat b_{2\perp} \! \right )^2\!-1\right ]
\nonumber\\
&&~~\times x_1 h_{  \gamma}\!(x_1,b_{1\perp}^2) x_2 h_{\gamma}\!(x_2,b_{2\perp}^2) \ . \label{ps2}
\end{eqnarray}
The above expressions are very similar to those in Ref.~\cite{Li:2019yzy}. We have also computed for the $v_2$ anisotropy, and found that they are power suppressed by the lepton mass in terms of $m_\ell/P_\perp$. This is because the $\cos(2\phi)$ asymmetry requires helicity flip in the QED process of $\gamma\gamma\to \ell^+\ell^-$, which is suppressed by the lepton mass. The lepton mass is too small to have any observational effects for $v_2$ for the typical kinematics at RHIC and LHC.

Note that there is no Sudakov effects in Eq.~(\ref{eq:e5}). This is because the Sudakov effects come from the incomplete cancellation between real and virtual contributions from higher-order QED corrections. Since in our current study, we integrate out small $q_\perp$, the real and virtual divergences cancel out completely. As a consequence, the Sudakov factor is absent in the differential cross section of Eq.~(\ref{eq:e5}).

\begin{figure}[htpb]
\includegraphics[angle=0,scale=0.8]{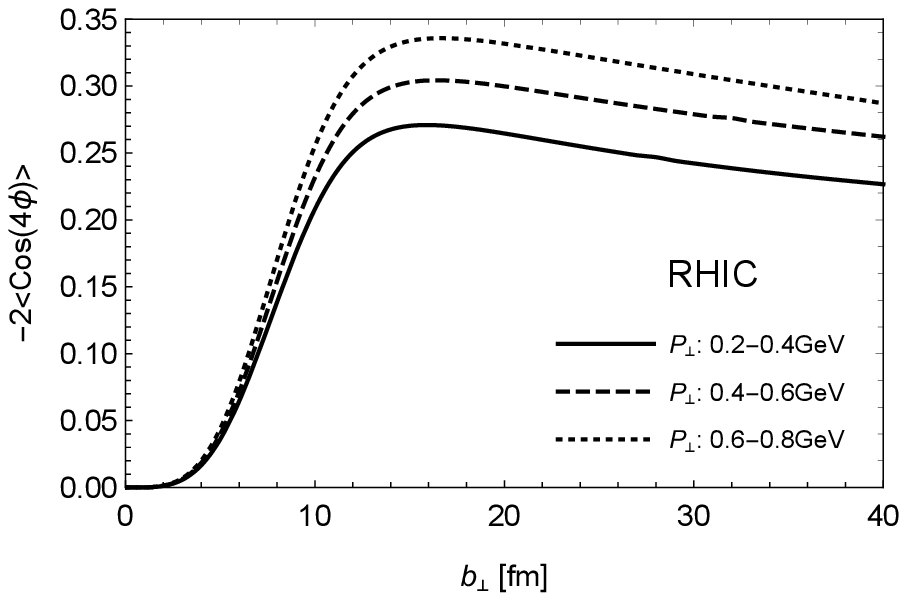}
\includegraphics[angle=0,scale=0.8]{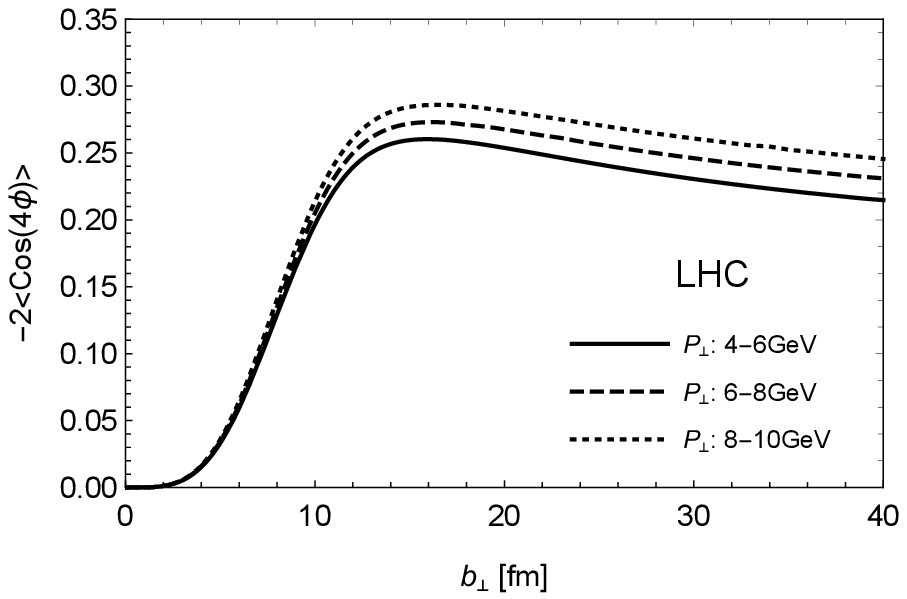}
\caption{ Estimates of the $\cos 4\phi$ asymmetry as the function of $b_\perp$
 in Au-Au collisions at  $\sqrt {s}=200 $ GeV(the upper plot) and in Pb-Pb collisions at  $\sqrt {s}=5.02 $ TeV(the lower plot). The dilepton rapidities are integrated over the regions [-1,1].
 } \label{fig2}
\end{figure}

\begin{figure}[htpb]
\includegraphics[angle=0,scale=0.8]{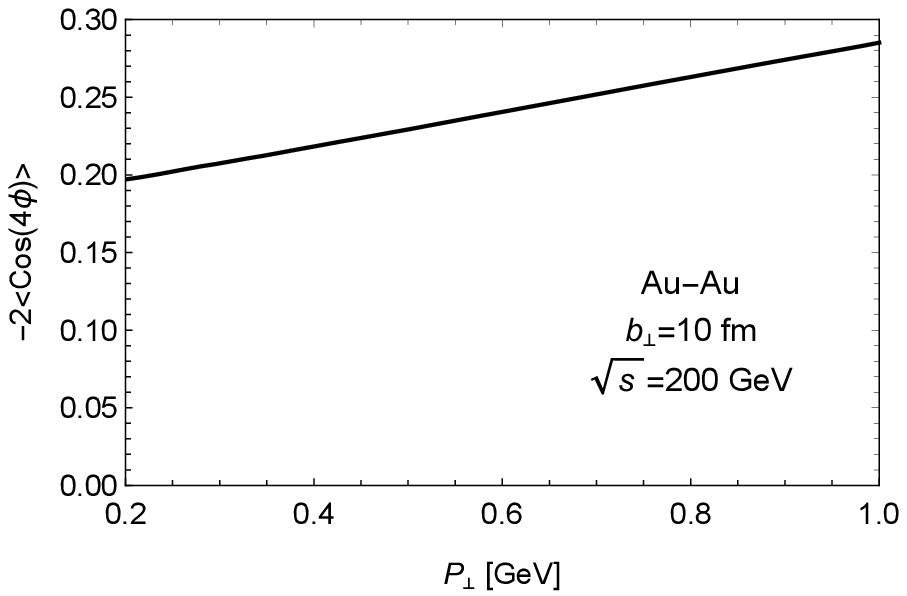}
\includegraphics[angle=0,scale=0.8]{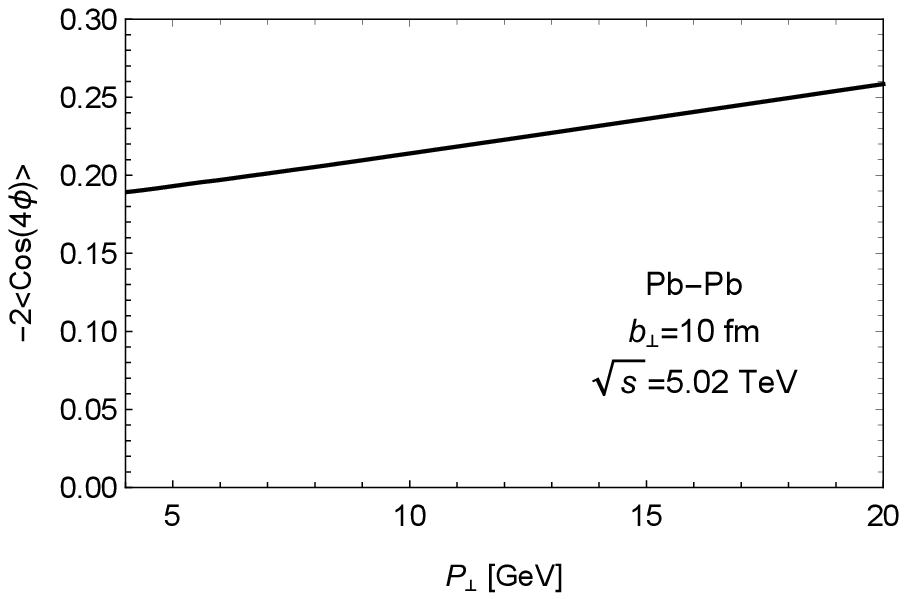}
\caption{ The asymmetries are plotted as the function of the lepton's transverse momentum.
 } \label{fig3}
\end{figure}

The nuclear charge form factor used in our numerical evaluation is taken from the STARlight MC generator~\cite{Klein:2016yzr},
\begin{eqnarray}
F_A(q^2)&=&\frac{4\pi \rho^0}{q^3 A}\frac{1}{a^2 q^2+1}\nonumber\\
&&\times \left [ \sin(qR_A)-qR_A \cos(qR_A)\right ]\ , \label{ff}
\end{eqnarray}
where $R_A=1.1 A^{1/3}$fm, and $a=0.7$fm. This parametrization numerically is very close to the Woods-Saxon distribution. The numerical results for the computed azimuthal asymmetries in the different kinematical regions for different collisions species are presented in Figs.~\ref{fig2} and \ref{fig3}. Here the azimuthal asymmetries, i.e. the average value of  $\cos (4\phi)$ are defined as,
\begin{eqnarray}
\langle \cos(4\phi) \rangle &=&\frac{ \int \frac{d \sigma}{d {\cal P.S.}} \cos (4\phi) \ d {\cal P.S.} } {\int \frac{d \sigma}{d {\cal P.S.}}  d {\cal P.S.}} \ .
\end{eqnarray}
In Fig.~\ref{fig2}, we show the asymmetries as the functions of the
impact parameter $b_\perp$ in the heavy-ion collisions at RHIC and LHC, respectively. As one can see, the general trend is that the asymmetry increases with $b_\perp$ until it reaches a maximal value when $b_\perp $ is slightly larger than $2R_A$. The maximal value of the asymmetry $-2\langle \cos(4\phi) \rangle$ ranges from  26\% to 34\% depending on the center-of-mass energy, lepton transverse momenta regions, and collision species. After reaching its maximal value, the asymmetry slowly decreases with $b_\perp$, but remain sizable untill the impact parameter is very large.

In Fig.~\ref{fig3}, we plot the asymmetries as functions of the lepton's transverse momentum $P_\perp$. Clearly, at both RHIC and LHC, the asymmetries do not change dramatically with $P_\perp$. This is very much different from hadron's $v_4$ in non-central heavy-ion collisions, where the transverse momentum dependence is one of the characteristic features of the medium flow. Of course, in the current case, we can not go to very small transverse momentum for the leptons. This is because we need to keep back-to-back kinematics ($q_\perp$ is small) for the dilepton and large transverse momentum for the leptons to guarantee the dominance from the QED two-photon scattering and the factorization formalism in Eqs.~(\ref{ps},\ref{ps2}). 

The momentum anisotropy of leptons can be measured through the azimuthal angular correlations between the lepton and hadrons, similar to what has been done for the hadron flow. Of course, due to the fluctuations, the fourth order event plane may not be well aligned with the impact parameter $b_\perp$. This will introduce considerable uncertainties in measuring the $v_4$ for leptons in central collisions, while this issue becomes less severe for peripheral events. The measurements will provide important information on the production mechanism for the dilepton.

\section{Conclusions}

In summary, we have studied the electromagnetic anisotropy of the leptons in heavy-ion collisions, where the leptons are produced in the pure QED process of $\gamma\gamma\to \ell^+\ell^-$. Our study shows that there is a significant size of $v_4$ anisotropy, whereas $v_2$ vanishes due to small lepton mass. These observables are defined as the azimuthal angular asymmetries of the lepton's transverse momentum with respect to the impact parameter of non-central heavy-ion collisions. The asymmetries evaluated in various kinematic regions and for different collisions species are shown to be rather sizable. The experiment confirmation of $v_4$-anisotropy in Figs.~\ref{fig2} and \ref{fig3} will help to identify the production mechanism of the lepton pair at low total transverse momentum in heavy-ion collisions. Any deviation will indicate other production channels. Once this is established, we can utilize the lepton pair to probe the EM property of the quark-gluon plasma.

The EM anisotropy of the leptons can be measured through the azimuthal angular correlations between the leptons and hadrons. The comparison of the phenomena between the lepton and hadrons shall provide useful information on the underlying physics in heavy ion collisions. In particular, because the lepton anisotropy comes from the initial EM field of the colliding nuclei in non-central collisions while the hadron flow comes from the collective modes in the quark-gluon plasma, a detailed study of both observables will lead to a deeper understanding of the interface of the initial geometry and later interactions of the medium. 

{\bf Acknowledgments:}  We thank Spencer Klein, Volker Koch and Zhangbu Xu for valuable comments and discussions. This material is based upon work supported by the U.S. Department of Energy, Office of Science, Office of Nuclear Physics, under contract number DE-AC02-05CH11231, and within the framework of the TMD Topical Collaboration. This work is also supported by the National Science Foundations of China under Grant No. 11675093, No. 11575070.

\end{document}